\documentclass[a4paper]{mem}
\usepackage{natbib}
\usepackage{graphicx}
\idline{75}{282}
\begin{document}

\title{
Near-Infrared Observations of RR Lyrae Variables \newline in $\omega$ Centauri
}

   \subtitle{}

\author{
M. \,Del Principe\inst{1}, A.M. \,Piersimoni\inst{1}, J. \,Storm\inst{2},
 G. \,Bono\inst{3}, F. \,Caputo\inst{3}, S. \,Cassisi\inst{1}, 
L.M. \,Freyhammer\inst{4}, M. \, Marconi\inst{5}, 
\and P.B. \, Stetson\inst{6}
          }

  \offprints{M. Del Principe}

\institute{
INAF-Osserv. Astron. di Collurania, via M. Maggini, 64100 Teramo, Italy
\and
Astrophysikalisches Institut Potsdam, Sternwarte 16, 14482 Potsdam, Germany
\and
INAF-Osserv. Astron. di Roma, via Frascati 33, 00040 Monte Porzio, Italy
\and 
Royal Observatory of Belgium, Ringlaan 3, 1180 Brussels, Belgium
\and
INAF-Osserv. Astron. di Capodimonte, via Moiariello 16, 80131 Napoli, Italy  
\and
Dominion Astrophysical Observatory, Herzberg Institute of Astrophysics,
National Research Council, 5071 West Saanich Road, Victoria, BC V9E~2E7
}

\authorrunning{Del Principe et al. }

\titlerunning{NIR Observations of RR Lyrae Variables}

\abstract{
We present Near-Infrared (NIR) $J$ and $K_s$-band observations for 181 RR Lyrae stars 
in the Galactic Globular Cluster $\omega$ Cen. The comparison between 
predicted and empirical slopes of NIR Period-Luminosity (PL) relations indicates 
a very good agreement. Cluster distance estimates based on NIR PL relations agree 
quite well with recent determinations based on different standard candles, giving  
a true mean distance modulus $\mu = 13.71\pm0.05$, and $d=5.52\pm0.13$ kpc.

}
\maketitle{}

\section{Introduction}

During the last few years, a substantial theoretical effort has been devoted to
the pulsational properties of RR Lyrae stars in NIR bands. These investigations
rely on predictions based either on pulsational models (Bono et al. 2001, 2003) 
or on synthetic horizontal branches (Catelan et al. 2004, Cat04; Cassisi et al.
2004, Cas04),
and provide PL relations in different NIR bands. 
However, observations lag theoretical predictions, and in order to fill this 
observational gap and to validate current predictions concerning NIR PL relations, 
we have undertaken a long-term project aimed at collecting NIR observations for 
RR Lyrae stars
in clusters in the Milkyway and in the Large Magellanic Cloud.             

\section{Discussion of results}

The $JK_s$ data for $\omega$ Cen were collected using the NIR camera 
SOFI@NTT.
We derived accurate $J$ and $K_s$ 
light curves for 56 RR Lyrae. Together with these data we extracted from the ESO 
archive other $J$ and $K_s$ images (Sollima et al. 2004) and we derived a 
few or single epoch $J$ and $K_s$ measurements for 58 additional RR Lyrae.
In order to further increase the sample of RR Lyrae stars we also added the 29 RR Lyrae 
observed by Longmore et al. (1990) and the 38 RR Lyrae with single epoch $J$ and 
$K_s$ magnitudes provided by the 2MASS survey. We ended up with a sample 
of 82 RR$ab$ and 99 RR$c$ with at least one $J$ or  
$K_s$ measurement, for a total of 181 out of the 
186 RR Lyrae in $\omega$ Cen with accurate periods.

The pre-reduction was performed using standard IRAF procedures and the photometry  
using DAOPHOT/ALLFRAME packages. The absolute calibration in the 2MASS system was 
performed using $\approx$400 local standard stars. The accuracy of the absolute 
zero-point is $\sim0.02$ mag. We fitted the individual $K_s$-band phase points with a template curve 
(Jones et al. 1996). This method requires for each variable accurate estimates 
of both the epoch and the luminosity amplitudes, which are only available for 
the RR Lyrae observed by Kaluzny et al. (2004). The mean $J$-band magnitude of 
well-sampled light curves were estimated with a spline fit.  
Figure 1 shows selected $J$ and $K_s$-band light curves for two first 
overtones (RR$c$) and two fundamental (RR$ab$) variables. 

To estimate the distance to $\omega$ Cen, we adopted the $K$-band PL relation 
provided by Cas04, a new $J$-band PL relation computed using the same evolutionary and
pulsation models of Cas04\footnote{For Z=0.001 and a HB type=0.90, i.e. the HB type of $\omega$
Centauri, we found: $\mathrm{<M_J>=-1.70(\pm0.03)LogP-0.624(\pm0.03)}$.}, 
and the $J$,$K$-band PL relations provided by Cat04, 
transformed into the 2MASS NIR system using the transformations provided by 
Carpenter et al. (2001). By assuming a constant reddening value 
$E(B-V)=0.11$ mag, we obtained a true distance modulus 
$\mu=(J-M_J)_0=13.66\pm0.07$ mag and $\mu=(K-M_K)_0=13.67\pm0.04$ mag
by adopting the new $J$-band PL relation 
and the $K$-band PL relation given by Cas04. 
Interestingly enough, the $J$ and $K$-band PL relations provided by 
Cat04 give respectively $\mathrm{\mu =13.75\pm0.05}$ mag and 
$\mathrm{\mu=13.73\pm0.04}$ mag. These estimates agree, within empirical 
and theoretical uncertainties,  with recent estimates based on different distance 
indicators, such as an eclipsing binary (Thompson et al. 2001). They are  
roughly 15\% longer than the dynamical distance estimates by  
van de Ven et al. (2005). 

\begin{figure}
\includegraphics[width=6.5cm, height=5.8cm]{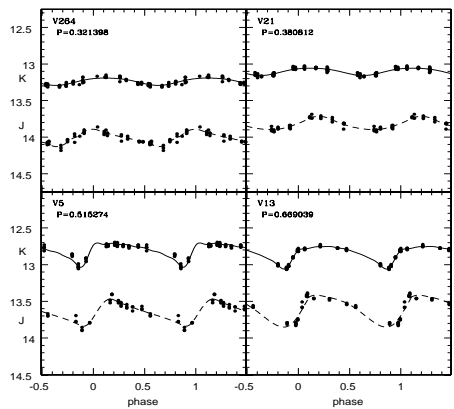}

\caption{Well-sampled light curves for two RR$c$ and two 
RR$ab$. Solid and dashed lines show the $K$-band template fit and the spline fit.} \label{var}
\end{figure}

\bibliographystyle{aa}

\end{document}